\documentstyle[12pt]{article}
\begin{document}
\title{Feynman-Kac Kernels in Markovian Representations of the
Schr\"{o}dinger Interpolating Dynamics}
\author{Piotr Garbaczewski  and Robert Olkiewicz\\
Institute of Theoretical Physics, University of Wroc{\l}aw,\\
PL-50 204 Wroc{\l}aw, Poland}
\maketitle
\hspace*{1cm}
\begin{abstract}
Probabilistic solutions of the so called Schr\"{o}dinger boundary data
problem provide for a unique Markovian interpolation between any two
strictly positive probability densities designed to form the
input-output statistics data for the process taking place in a
finite-time interval. The key issue is to select the jointly
continuous in all variables positive Feynman-Kac kernel, appropriate
for the phenomenological (physical) situation.
We extend the existing formulations of the problem
to cases when the kernel is \it not \rm a fundamental solution of a
parabolic equation, and prove the existence of a continuous Markov
interpolation in this case. Next, we analyze the compatibility of
this stochastic evolution with the original parabolic dynamics,
while assumed to be  governed
by the temporally adjoint pair of (parabolic) partial differential
equations, and prove that the pertinent random motion  is a
diffusion process.  In particular, in conjunction with Born's
statistical interpretation postulate in quantum theory, we consider
stochastic processes which are compatible with the Schr\"{o}dinger
picture quantum evolution.
\end{abstract}

\section{Motivation: Schr\"{o}dinger's interpolation problem through
Feynman-Kac kernels}

The issue of \it deriving \rm a microscopic dynamics  
from the  (phenomenologically  or numerically motivated, by
approximating the frequency distributions) input-output statistics
data was addressed,
as the Schr\"{o}dinger problem of a probabilistic interpolation,
in a number of publications
\cite{schr}-\cite{olk1}.
We shall consider Markovian propagation scenarios so remaining within
the well established framework, where for any two Borel sets
$A,B\subset R$ on which
the respective strictly positive boundary  densities $\rho (x,0)$  
and $\rho (x,T)$ are defined, the transition probability
$m(A,B)$ from the set $A$ to the set $B$ in
the time interval $T>0$ has a density given in a specific factorized
form: 
$$m(x,y)=f(x)k(x,0,y,T)g(y)$$
$$m(A,B)=\int_Adx\int_Bdy \, m(x,y)$$
$${\int dy m(x,y)=\rho (x,0)\, ,\, \int dx m(x,y)=\rho (y,T)}
\eqno (1)$$

Here, $f(x), g(y)$ are the a priori unknown functions, to come out as 
solutions of the integral (Schr\"{o}dinger) system of equations (1),
provided that in addition to the density boundary data we have in
hands 
any strictly positive, continuous  in space variables \it function \rm 
$k(x,0,y,T)$. 
Our notation makes explicit the dependence (in general irrelevant) 
 on the time interval endpoints.  It anticipates an important
restriction we shall impose, that $k(x,0,y,T)$ must be  a strongly 
continuous dynamical semigroup  kernel:  it will secure  the 
Markov property of the sought for stochastic process.

It is the major mathematical discovery \cite{jam} that,
without the semigroup 
assumption \it but \rm with the prescribed, nonzero boundary data 
$\rho (x,0),\rho (y,T)$ \it and \rm with the  strictly positive
continuous function
$k(y,0,x,T)$, the Schr\"{o}dinger system (1) of integral equations
admits a unique solution in terms of two nonzero, locally integrable 
 functions $f(x), g(y)$ of the same
sign (positive, everything is up to a multiplicative constant).

If $k(y,0,x,T)$ is a particular, confined to the time interval
endpoints,  form of a concrete semigroup kernel
$k(y,s,x,t), 0\leq s\leq t<T$, let
it be a fundamental solution associated with (5) (whose existence 
a priori is \it not \rm granted),  
then there exists 
\cite{zambr,blanch,olk,olk1,nag} a function $p(y,s,x,t)$:
$${p(y,s,x,t)=k(y,s,x,t){{\theta (x,t)}\over {\theta (y,s)}}}
\eqno (2)$$
where 
$${\theta (x,t)=\int dy k(x,t,y,T)g(y)}\eqno (3)$$
$$\theta _*(y,s)=\int dx k(x,0,y,s)f(x)$$
which implements a consistent propagation of the density 
$\rho (x,t)=\theta (x,t)\theta _*(x,t)$ between its boundary versions,  
according to:
$${\rho (x,t) = \int p(y,s,x,t)\rho (y,s)dy}\eqno (4)$$
$$0\leq s\leq t<T$$
 For a given semigroup which is characterized by its generator
 (Hamiltonian), the kernel $k(y,s,x,t)$ and the emerging transition
 probability density
$p(y,s,x,t)$  are unique in view of the uniqueness of solutions
$f(x),g(y)$  of (1). For Markov processes, the knowledge of the
transition probability density $p(y,s,x,t)$ for all intermediate
times $0\leq s< t\leq T$  suffices
for the derivation of all other relevant characteristics. 

In the framework of the Schr\"{o}dinger problem the choice of the
integral  kernel $k(y,0,x,T)$ is arbitrary,
except for the strict positivity and
continuity demand. As long as there is no "natural" 
physical motivation for its concrete functional form, the problem is  
abstract and of no direct physical relevance. 

However, in the context of parabolic partial differential equations
this "natural" choice is automatically settled if the Feynman-Kac
formula can be utilized to represent solutions.
Indeed, in this case an unambigous strictly positive
semigroup kernel which is a continuous function of its arguments,
can be 
introduced for a broad class of (admissible \cite{simon})
potentials. Time 
dependent potentials are here included as well \cite{freid,simon1}.
Moreover, in Ref. \cite{blanch} we have discussed a possible 
phenomenological significance  of the Feynman-Kac potentials,
as contrasted  
to the usual identification of Smoluchowski drifts with force fields 
affecting particles (up to a coefficient) in the standard theory of 
stochastic diffusion  processes.

In the existing probabilistic investigations
\cite{zambr,zambr1,garb,blanch,olk}, based on the exploitation of the 
Schr\"{o}dinger problem strategy, it was generally assumed that
the kernel actually  \it is  \rm  a  fundamental solution of the
parabolic equation.
It means that the kernel is a function with continuous 
derivatives: first order-with respect to time, second order-with
respect to space variables.
Then, the transition probability density defined by (2) is a
fundamental solution of the Fokker-Planck (second Kolmogorov)
 equation in the pair $x,t$ of variables, and as such is at the
 same time a solution of the
backward (first Kolmogorov) equation in the pair $y,s$.
This feature was  exploited in \cite{blanch,olk}.

There is a number of mathematical subtleties involved in the
fundamental solution notion,
since in this case, the Feynman-Kac kernel must be  a
solution of  the parabolic equation itself.
In general, Feynman-Kac kernels may
have granted the existence status, even as continuous functions
\cite{simon,simon1,glimm}, but may not  be differentiable,
and need not to be solutions of any conceivable partial differential 
equations. 

To our knowledge, this complication in the study of Markovian 
representations of the Schr\"{o}dinger  interpolating dynamics
(and the quantum Schr\"{o}dinger picture dynamics in particular)
has never been addressed in the literature. Moreover, it is far from
being obvious that this Markovian interpolation
 actually \it is \rm a diffusion process.

\section{Schr\"{o}dinger's interpolation problem: general
derivation of the stochastic evolution}

\subsection{The Schr\"{o}dinger system of integral equations}

We shall complement our previous analysis \cite{blanch,olk} by 
discussing the issue in more detail. It turns out the the crucial step
lies in a \it proper \rm choice of the strictly positive and
continuous 
function $k(y,s,x,t), s<t$ which, if we want to construct a Markov 
process, has to satisfy the Chapman-Kolmogorov (semigroup composition) 
equation. To proceed generally let us
consider a pair of partial differential equations for real 
functions $u(x,t)$ and $v(x,t)$:
$${\partial _tu(x,t)=\triangle u(x,t) - c(x,t)u(x,t)}\eqno (5)$$
$$\partial _tv(x,t)= -\triangle v(x,t) + c(x,t)v(x,t)$$
where,  we have eliminated all
unnecessary  dimensional parameters.

Usually, \cite{glimm,simon}, $c(x,t)$ is assumed to be a 
continuous and bounded from below function.
We shall adopt weaker  conditions.
Namely, let us decompose $c(x,t)$ into a sum of positive and
negative terms:
$c(x,t)=c_+(x,t) - c_-(x,t)\; ,\; c_{\pm }\geq 0$ where 
(a) $c_-(x,t)$ is bounded,  while (b) $c_+(x,t)$ is bounded on compact 
sets of $R\times [0,T]$. It means that $c(x,t)$ is bounded from below
and 
locally  bounded from above. Clearly,  $c(x,t)$
needs not to be a continuous  function and then we  encounter
weak solutions of (5) which admit discontinuities.

With the first (forward) equation (5)  we can  immediately
associate an  integral kernel of the time-dependent semigroup
(the exponential operator should be understood as the time-ordered
expression):
$${k(y,s,x,t)=[exp(-\int_s^t H(\tau )d\tau )](y,x)}\eqno (6)$$
where $H(\tau )=-\triangle +c(\tau )$. 
It is clear, that for discontinuous $c(x,t)$, no fundamental
solutions are  admitted by (5).

By the Feynman-Kac formula, \cite{simon1,freid}, we get
$${k(y,s,x,t)=\int exp[-\int_s^tc(\omega (\tau ),\tau)d\tau ]
d\mu ^{(y,s)}_{(x,t)}(\omega )}\eqno (7) $$
where $d\mu ^{(y,s)}_{(x,t)}(\omega)$ is the conditional Wiener 
measure over sample paths of the standard Brownian motion.

It is well known that $k$ is strictly positive in case of $c(x,t)$ 
which is  continuous and bounded from below; typical proofs are given 
 under an additional assumption that $c$ does not depend on time 
\cite{glimm}. However, our assumptions  about $c(x,t)$ were weaker, 
and to see that nonetheless  $k$ is strictly positive we shall follow 
the  idea of Theorem 3.3.3 in \cite{glimm}.  Namely, the conditional
Wiener measure $d\mu _{(x,t)}^{(y,s)}$ can be written as follows
$${
d\mu _{(x,t)}^{(y,s)} = [4\pi (t-s)]^{-1/2} 
exp[-{{(x-y)^2}\over {4(t-s)}}]\: d\nu _{(x,t)}^{(y,s)}}\eqno (8)$$
where $d\nu _{(x,t)}^{(y,s)}$ is the normalised Wiener measure 
\cite{simon}. We can always choose a certain  number $r>0$ to
constrain  the event (sample path) set
$${
\Omega (r)=[\omega : X_s(\omega )=y, X_t(\omega )=x,
sup_{s\leq \tau \leq t}\: |X_{\tau }(\omega )|\leq r]}\eqno (9)$$
It comprises these sample trajectories which are bounded by $r$ on the
time interval $[s,t]$. In the above,  $X_t(\omega )$ is the value
taken by 
the random variable $X(t)$ at time $t$, while  a concrete
$\omega $-th 
path is sampled. By properly tuning $r$, we can always achieve
$${
\int_{\Omega (r)} d\nu _{(x,t)}^{(y,s)} \geq {1\over 2}}\eqno (10)$$
which  implies that
$${
k(y,s,x,t)\geq {1\over 2} [4\pi (t-s)]^{-1/2}\: 
exp[-{{(x-y)^2}\over {4(t-s)}}]\: exp[-(t-s)C] > 0}\eqno (11)$$
$$C=sup_{s\leq \tau \leq t,\; \omega \in 
\Omega (r)}\;  c_+(X_{\tau }(\omega ),\tau )$$
where, by our assumptions, $c_+$ is bounded on compact sets.
Consequently, the kernel $k$ is \it strictly positive \rm .

With the Schr\"{o}dinger boundary data problem on mind, we must settle
an   issue of the \it continuity \rm of the kernel.
To this end, let us invoke
 a well known procedure  of rescaling of path  integrals 
\cite{simon,roep}:
by passing from the "unscaled" sample paths $\omega (t)$ 
over which the conditional Wiener measure integrates,
to the "scaled" paths 
of the Brownian bridge, the $(y,x)$ conditioning can be 
taken away from the measure. 
Then, instead of sample paths $\omega $ 
 connecting points $y$ and $x$ in the time interval $t-s>0$, 
we consider the  appropriately "scaled" paths  of the Brownian bridge 
$\alpha  $ connecting the point $0$ with $0$ again, in the (scaled)
time $1$. 
It is possible, in view of the  decomposition \cite{simon,roep}:
$${\omega (\tau )=({t\over {t-s}}- {\tau \over {t-s}})y  +  
({\tau \over {t-s}} - {s\over {t-s}})x + \sqrt{t-s} \; 
\alpha ({\tau \over {t-s}} - {s\over {t-s}})}\eqno (12)$$
where $\alpha $ stands for the "scaled" Brownian bridge. Then, 
we can write 
$${k(y,s,x,t)=[4\pi (t-s)]^{-1/2} exp[-{(x-y)^2\over {4(t-s)}}] 
\int d\mu (\alpha )\cdot }\eqno (13)$$
$$exp[- \int_s^t c({{t-\tau }\over {t-s}}y + 
{{\tau -s}\over {t-s}}x + \sqrt{t-s}\: \alpha
({{\tau -s}\over {t-s}})\; 
, \tau )d\tau ]$$
where $d\mu (\alpha )=d\nu ^{(0,0)}_{(0,1)}(\omega )$ is the
normalized  Wiener  measure
integrating with respect to the "scaled" Brownian bridge paths, which 
begin and terminate at the origin $0$ in-between "scaled time"
instants: $0$ corresponding to $\tau =s$ and $1$ corresponding to
$\tau =t$.

This representation of   $k$, \it if \rm combined with the  assumption
that $c(x,t)$ is a continuous function, allows to conclude,
\cite{simon},
that the kernel is continuous in all variables. However, our previous
assumptions were weaker,
and it is instructive to know that through suitable approximation
techniques,  Theorem B.7.1 in Ref.\cite{simon1} proves that the
kernel is jointly continuous in our case as well.

It is also clear that $k(y,s,x,t)$
satisfies the Chapman-Kolmogorov composition rule.
So, the first equation (5) can be used to define the  Feynman-Kac
kernel, appropriate for the Schr\"{o}dinger problem analysis
in terms of a Markov stochastic process.

Let us consider an arbitrary (at the moment) pair of strictly
positive,  but not necessarily continuous, boundary densities
$\rho _0(x)$ and $\rho _T(x)$. By Jamison's
principal theorem \cite{jam} there exists a unique pair of strictly 
positive, locally (i.e. on compact sets) integrable functions $f(x)$ 
and $g(x)$ solving the Schr\"{o}dinger system (1), e.g. such that
${\rho _0(x)=f(x)\int k(x,0,y,T)g(y)dy}$ and 
$\rho _T(x)=g(x)\int k(y,0,x,T)f(y)dy$
with the kernel $k(y,s,x,t)$ given by (7).

Let us define:
$${g(x,t)=\int k(x,t,y,T)g(y)dy\; \;  ,\; \;  f(x,t)=
\int k(y,0,x,t)f(y)dy}
\eqno (14)$$
The above integrals exist at least for almost every $x$
so that there appears
the problem of the existence of a unique and continuous transition
probability density  $p(y,s,x,t)$, (2).
We shall assume that
the function $g(y)$ is bounded at infinity. This means that there
exists a constant $C>0$ and a compact set $K\subset R$ such that
$g(y)\leq C$ for all $y\in R\backslash K$.
Then, for all $t<T$ and any sequences
$h_n\rightarrow 0 , s_n\rightarrow 0$, as $n\rightarrow \infty
$ we get ($lim$ stands for $lim_{n\rightarrow \infty }$):
$$lim \: |g(x+h_n,t+s_n)- g(x,t)|  \leq lim\: |\int_K
[k(x+h_n,t+s_n,y,T)-k(x,t,y,T)]g(y)dy| \: +\: $$
$${lim\: |\int_{R\backslash K} [k(x+h_n,t+s_n,y,T)-
k(x,t,y,T)]g(y)dy|\leq }\eqno (15)$$
$$lim\: sup_{y\in K}\:
|k(x+h_n,t+s_n,y,T)-k(x,t,y,T)|\int_K g(y)dy \: +\: $$
$$C\cdot  lim\: \int_{R\backslash K}
|k(x+h_n,t+s_n,y,T)-k(x,t,y,T)|dy$$
The first term tends to zero because $k$ is jointly continuous and $g$
is locally integrable.The second one  tends to zero because of the
Lebesgue bounded convergence theorem.
Consequently, our assumption
suffices to make $g(x,t)$ continuous on $R\times [0,T)$.
Similarly, we can prove that $g(x,t)$ is bounded. \\

Now, we can  set according to (2),
$p(y,s,x,t)=k(y,s,x,t)g(x,t)/g(y,s)$.
Then,  
$p(y,s,x,t)$, $0\leq s<t\leq T$ becomes a transition
probability density
of a Markov stochastic process with a factorized density 
$\rho (x,t)=f(x,t)g(x,t)$. 
Clearly, this stochastic process  interpolates between the
boundary data
$\rho _0$ and $\rho _T$ as time continuously varies from $0$ to $T$.
Notice that (15) implies the continuity of $p$ in the
time interval $[0,T)$.

Although $p(y,s,x,t)$ is continuous in all variables, we cannot be
sure that the interpolating stochastic process has  continuous
trajectories,  and
no specific (e.g. Fokker-Planck) partial differential 
equation can be readily  associated with 
this dynamics.
Therefore, we must explicitly verify whether the
associated process is  stochastically continuous. If so, we should
know  whether it is continuous (i.e. admits continuous trajectories).
Eventually, we should  check the vailidity of  conditions under
which the investigated interpolation can be regarded as
 a diffusion process. The subsequent analysis will prove that this
ultimate goal results only due  to the gradual strengthening of
conditions imposed on the parabolic system (5).
 
\subsection{Stochastic continuity of the process}

Apart from the  generality of formulation of the 
Schr\"{o}dinger interpolation problem  which appears to
preclude an unambigous identification (diffusion or not) of the 
constructed stochastic process, we can prove in the present case, 
a fundamental property of a
stochastic dynamics called a stochastic 
continuity of the process. In this connection, compare e.g.
\cite{zambr,dynk} and \cite{nel1}, where this 
property is linked to the uniqueness of the corresponding Markov
semigroup  generator. The stochastic continuity property is
a necessary condition for the process to admit
continuous trajectories.

The stochastic process is stochastically continuous, if  for  the 
probability of the occurence of sample paths $\omega $,  
such that the random variable values $X_t(\omega )$ along the
trajectory 
obey $|X_t(\omega )-X_s(\omega )|\geq \epsilon \; ,\; s< t$,
the following limiting behaviour is recovered 
$${lim_{t\downarrow s} P[\omega:|X_t(\omega )-X_s(\omega )|
\geq \epsilon ]  = 0 }\eqno (16)$$
for every positive $\epsilon $. This demand can be written in a more
handy  way in terms of the transition probability density
$p(y,s,x,t)$ and the density $\rho (x,t)$ of the process:
$${
lim_{t\downarrow s} [\int_{-\infty }^{+\infty }dy \rho (y,s) 
\int_{|x-y|\geq \epsilon } p(y,s,x,t) dx ] = 0}\eqno (17)$$
So, for the transition density to be stochastically continuous, 
it suffices that
$${
lim_{\triangle s\downarrow 0} \int_{|x-y|\geq \epsilon } 
p(y,s,x,s+\triangle s)dx = 0}\eqno (18)$$
for almost every $y\in R$.

In view of our construction, (2), we have:
$${
lim_{\triangle s\downarrow 0} \int_{|x-y|\geq \epsilon } 
p(y,s,x,s+\triangle s)dx=}\eqno (19)$$
$${1\over {g(y,s)}} lim_{\triangle s\downarrow 0}
\int_{|x-y|\geq \epsilon }
dx \; k(y,s,x,s+\triangle s)\int_{-\infty }^{+\infty }  
k(x,s+\triangle s,z,T)\; g(z) dz$$
By changing the order of integrations (allowed by  positivity of the
involved functions)
we get:
$${
lim_{\triangle s\downarrow 0} \int_{|x-y|\geq \epsilon } 
p(y,s,x,s+\triangle s) dx = {1\over {g(y,s)}} lim_{\triangle s
\downarrow 0}
\int_{-\infty }^{+\infty } dz\: g(z)\: }\eqno (20)$$
$$[\int_{|x-y|\geq 0} dx \: k(y,s,x,s+\triangle s)\: 
k(x,s+\triangle s,z,T)]$$
Because the potential is bounded from below, 
$c\geq -M$ for some $M>0$, we 
easily arrive at the estimates (use the "scaled" Brownian bridge 
argument)
$${
k(y,s,x,s+\triangle s) \leq (4\pi \triangle s)^{-1/2}
exp[-{{(x-y)^2}\over {4\triangle s}}]
\: exp(M\triangle s)}\eqno (21) $$
and
$${
k(x,s+\triangle s,z,T)\leq [4\pi (T-s-\triangle s)]^{-1/2} \: 
exp[-{{(z-x)^2}\over {4(T-s-\triangle s)}}]}
{exp[M(T-s-\triangle s)]} \eqno (22)$$
Then we get:
$${
0\leq lim_{\triangle s\downarrow 0} \int_{|x-y|\geq \epsilon } 
k(y,s,x,s+\triangle s) k(x,s+\triangle s,z,T) dx  \leq  }$$
$${
[4\pi (T-s)]^{-1/2} exp[M(T-s)]
 lim_{\triangle s\downarrow 0}\: (4\pi \triangle s)^{-1/2}
 }\eqno (23)$$
$$
\int_{|x-y|\geq \epsilon } dx
exp[-{{(x-y)^2}\over {4\triangle s}}]\:
exp[-{{(z-x)^2}\over {4(T-s-\triangle s)}}] = 0$$
So, by the classic Lebesgue bounded (dominated) convergence theorem, 
the required limiting property 
$lim_{\triangle s\rightarrow 0} \int_{|x-y|\geq \epsilon } 
p(y,s,x,s+\triangle s) dx = 0 $ follows  and (16) holds true.

As mentioned before, the stochastic continuity  of the Markov process
is  a necessary condition
for the process to be continuous in a more pedestrian sense, i. e. to 
admit continuous sample paths. 
However, it is insufficient.  Hence, additional 
requirements are necessary to allow for a  standard diffusion 
process realization of solutions of  the general Schr\"{o}dinger 
problem, (1)-(3).
 
In the next section we shall prove that our process
can be regarded as continuous, by requiring a certain
correlation between the kernel $k(y,s,x,t)$ and a function
$g(x,t)$, (14).

\subsection{Continuity of the process}

It is well known that a solution of a parabolic equation cannot tend
to zero arbitrarily fast, when $|x|\rightarrow \infty $,
\cite{watson1}.
Roughly speaking, it cannot fall off faster than a fundamental
solution (provided it exists).
In fact, the solution is known to fall off as
fast as the fundamental solution, when the initial boundary data
coincide with the Dirac measure. If a support of the initial data is
spread (i.e. not point-wise), then the solution  falloff is slowlier
than this of the fundamental one.

In our discussion, where  $g(x,t)$ is a generalized solution and
$k(y,s,x,t)$ is a Feynman-Kac kernel which  does not need to be
a fundamental solution, we expect a similar behaviour.
Mathematically, our demand will be expressed as follows. Let
$t-s$ be small and $K$ be a compact subset in $R$.
Because $g(x,t)$ is supported on the whole $R$, so in the
decomposition
$${g(y,s)= \int_Kk(y,s,x,t)g(x,t)dx +
\int_{R\backslash K}k(y,s,x,t)g(x,t)dx}\eqno (24)$$
the second term  becomes relevant when $|y|\rightarrow \infty $ .
It amounts to (in the denominator there appears $g(y,s)$):
$${lim_{|y|\rightarrow \infty }\: {{\int_{-\infty }^{+\infty
}k(y,s,x,t)g(x,t)\chi _K(x)dx}\over {\int_{-\infty }^{+\infty }
k(y,s,x,t)g(x,t)dx}}\; = \; 0}\eqno (25) $$
where $\chi _K$ is an indicator function of the set $K$,
which is equal one for $x\in K$ and zero otherwise.

By means of the transition probability density $p(y,s,x,t)$ let us
introduce a transformation
$${(T_s^tf)(y) = \int_{-\infty }^{+\infty }p(y,s,x,t)f(x)dx}
\eqno (26)$$
of a function $f(x)$, continuous and vanishing  at infinity
(we shall use an abbreviation $f\in C_{\infty }(R)$ to express this
fact). It is clear that $(T_s^tf)(x)$ is a continuous function.
For a suitable  compact set $K$ we can always guarrantee the property
$|f(x)|<\epsilon $ for every $x\in R\backslash K$. Then, if we exploit
the property $\int_{R\backslash K}p(y,s,x,t)dx\leq 1$ if $s<t$ and the
definition of $p$ in terms of $k$ and $g$, we arrive at
$${|(T_s^tf)(y)|\; \leq \;
\int_Kp(y,s,x,t) |f(x,t)|dx\: +\: \int_{R\backslash K}
p(y,s,x,t)|f(x,t)|dx\; \leq }\eqno (27)$$
$$[\int_Kp(y,s,x,t)dx]\: \int_K|f(x,t)|dx\: +\:
sup_{x\in R\backslash K}\;
|f(x,t)|\: \int_{R\backslash K} p(y,s,x,t)dx \: \leq $$
$$[\int_K|f(x,t)|dx]\: {{\int_Kk(y,s,x,t)g(x,t)dx}\over {\int_{-\infty
}^{+\infty }k(y,s,x,t)g(x,t)dx}}\; +\; \epsilon $$
It implies that for small $t-s$, $lim_{|y|\rightarrow \infty }
(T_s^tf)(y)=0$, and so
$T_s^t$ forms an inhomogeneous in time semigroup of positive
contractions on $C_{\infty }(R)$.  For arbitrary $t$ and $s$ the
result follows by the obvious decomposition property
$T_s^t=T_s^{s_1}T_{s_1}^{s_2}\cdot \cdot \cdot T_{s_n}^t$.
In the well established terminology,
our $p(y,s,x,t)$ is a $C_{\infty }$-Feller transition function and
leads to a regular Markov process, \cite{dynk}.
Moreover, by the stochastic
continuity of $p(y,s,x,t)$, $T_s^t$ is strongly continuous.

As yet, we do not know whether  the process itself is continuous i.e.
has continuous sample paths. To this end,
it suffices to check whether the so called "Dynkin
condition", \cite{karlin}
$${ lim_{t\downarrow s}{1\over {t-s}}\: sup_{y\in K}\; [\int_{|x-y|>
\epsilon } p(y,s,x,t)dx]\: =\: 0}\eqno (28)$$
is valid for every $\epsilon >0$ and every compact set $K$. We have
(remember that $g(x,t)$ is strictly positive, continuous  and  bounded):
$$sup_{y\in K}\: \int_{|x-y|> \epsilon } p(y,s,x,t)dx\; =\; sup_{y
\in K}\: {1\over {g(y,s)}}\int_{|x-y|> \epsilon }k(y,s,x,t)g(x,t)dx\:
\leq $$
$${{{sup_{x}\: g(x,t)}\over {inf_{y\in K}\: g(y,s)}}\: \int_{|x-y|>
\epsilon }k(y,s,x,t)dx\: \leq \: C\: \int_{|x-y|> \epsilon
}k_0(x-y,t-s)dx}\eqno (29)$$
where (compare e.g. the previous estimate (22))
$${C={{sup_{x}\: g(x,t)}\over {inf_{y\in K}\: g(y,s)}}\;
exp[M(t-s)]}\eqno (30)$$
and $k_0(x-y,t-s)$ is the heat kernel.

Finally, we arrive at:
$${lim_{t\downarrow s} {1\over {t-s}}\: sup_{y\in K}\; [\int_{|x-y|>
\epsilon } p(y,s,x,t)dx]\; \leq }$$
$${ C\: lim_{t\downarrow s}{1\over
{t-s}}\int_{|z|> \epsilon }k_0(z,t-s)dz\: =\: 0}\eqno (31)$$
So, the stochastic process we are dealing with, is continuous.
Interestingly, "a continuous in
time parameter stochastic processes which possesses the (strong)
Markov property and for which  the sample paths $X(t)$ are
almost always (i.e.
with probability one) continuous functions of $t$ is called a
diffusion process", see e.g. chapter 15 of \cite{karlin}.

\subsection{The interpolating stochastic dynamics: compatibility  with
the temporally  adjoint  parabolic evolutions}

The  formulas (14) determine what is called, \cite{freid}, the
generalized solution of a parabolic equation: it  admits functions
which are
not necessarily continuous and if continuous, then not necessarily 
differentiable.  Before, we have established  the continuity of the
generalized solution $g(x,t)$ under rather mild  assumption about the
behaviour of $g(x)$ at spatial infinity. In fact, the same assumption
works for $f(x,t)$.
But nothing  has been said about the differentiability
of $f(x,t)$ and $g(x,t)$.

Consequently, our reasoning seems to be somewhat divorced from the
original partial  differential equations (5),
for which we can take for granted that certain
solutions $u(x,t)$ and $v(x,t)$ exist in the time interval
$0\leq t\leq T$. 
For this, we must assume that $c(x,t)$ is a continuous function.

Let us  consider the solutions of (5) that  are bounded functions of
their arguments. It is instructive to point out that we do not impose
any restrictions on the growth of $c(x,t)$ when
$|x|\rightarrow \infty $, and consequently we do not assume that
solutions of parabolic equations (5) have bounded derivatives.
Then, \cite{freid}, the solution $u(x,t)$ of the forward 
parabolic equation (5) is known to admit the Feynman-Kac representation
with the integral kernel (7),(13),  where
$${u(x,t)=\int k(y,s,x,t)u(y,s)dy}\eqno (32)$$
for $0\leq s<t\leq T$.
At this point let us define 
$${U(x,t)=v(x,T-t)}\eqno (33)$$
for all $t\in [0,T]$ and observe 
that, as a consequence of the time adjoint equation (5) for which
$v(x,t)$ \it is \rm a solution, the newly 
introduced function $U(x,t)$ solves the forward equation (5):
$${\partial _tU(x,t)=\triangle U(x,t) - c(x,T-t)U(x,t)}\eqno (34)$$
with a slightly rearranged potential: $c(x,t)\rightarrow  c(x,T-t)$.
By the assumed boundedness of the solution $v(x,t)$ of (5), we
arrive at the  Feynman-Kac formula 
$${U(x,t)=\int K(y,s,x,t)U(y,s)dy}\eqno (35)$$
with the corresponding kernel $K(y,s,x,t)$ of the  (time ordering
implicit)
operator $exp[-\int_s^tH(T-\tau )d\tau ]$, where 
$H(T-\tau )=-\triangle + c(T-\tau )$.
Let us emphasize  that in case of the 
time independent potential, $c(x,t)=c(x)$ for all $0\leq t\leq T$,
the  kernel $K$ coincides with $k$.

The previous Brownian bridge argument (12), (13) retains its validity,
and we have: 
$${K(y,s,x,t)=[4\pi (t-s)]^{-1/2} exp[-{{(x-y)^2}\over {4(t-s)}}]\cdot }
\eqno (36)$$
$$\int d\mu (\alpha )\:  exp[-\int_s^t c({{t-\tau }\over {t-s}}y 
+ {{\tau -s}\over {t-s}}x + \sqrt{t-s}\: 
\alpha ({{\tau -s}\over {t-s}})\: ,T-\tau )d\tau ]$$
which, after specializing  to the case of $s=0,t=T$  and accounting
for the  invariance of the Brownian bridge measure
with respect to the replacement
of sample paths  $\omega (\tau )$ by sample paths $\omega (T-\tau )$,
   \cite{nag,nel}, gives rise to:
$${K(y,0,x,T)=(4\pi T)^{-1/2} exp[-{{(x-y)^2}\over {4T}}]\cdot }\eqno (37)$$
$$\int d\mu (\alpha )\: exp[-\int_0^T c({\sigma \over T}y + (1-
{\sigma \over T})x +
 \sqrt{T}\: \alpha ({\sigma \over T}), \sigma )d\sigma ]$$
where $\sigma =T-\tau $.\\

A comparison of (37) with (13) proves that  we  have
derived an identity:
$${K(y,0,x,T)=k(x,0,y,T)}\eqno (38)$$
whose immediate consequence is the formula 
$${U(x,T)=v(x,0)=\int k(x,0,y,T)v(y,T)dy}\eqno (39)$$
for the backward propagation of $v(y,T)$ into $v(x,0)$.

We shall utilize (39) and (32), under an \it additional \rm
assumption
that the previous, hitherto arbitrary, probability density data 
$\rho _0(x), \rho _T(x)$, actually are 
determined by the initial and terminal values of 
the  solutions $u(x,t),\: v(x,t)$  of (5), according to:
$${\rho _0(x)=u(x,0)v(x,0)}$$
$${\rho _T(x)=u(x,T)v(x,T)}\eqno (40)$$
Our present aim is to show that with this assumption, we can identify
the (still abstract) functions $f(x,t)$, $ g(x,t)$, (14), with $u(x,t)$
and $v(x,t)$ respectively.
By (32), (39) there holds:
$${\rho _0(x)=u(x,0)\int k(x,0,y,T) v(y,T)dy}\eqno (41)$$
$$\rho _T(x)=v(x,T)\int k(y,0,x,T)u(y,0)dy$$
and, in view of the uniqueness of solution of the Schr\"{o}dinger
system, once the boundary densities and the continuous strictly
positive kernel are specified,
we realize that the propagation formulas (14)
involve solutions of (5) through the respectively initial
and terminal data:
$$ {f(x)=u(x,0)}$$
$${g(x)=v(x,T)}\eqno (42)$$
Moreover, (5),(14) imply that $f(x,t)=u(x,t)$ holds true identically
for  all $t\in [0,T]$. 

What remains to be settled is whether the function  $g(x,t)$ can be 
identified with the solution $v(x,t)$ of (5) for all $t\in [0,T]$.

This property is obvious, when the
time independent potential $c(x)$ is investigated instead of the  more 
general $c(x,t)$. As well, the identification is with no doubt in case 
when $k(y,s,x,t)$ is a fundamental solution of the parabolic equation  
in variables $x,t$. In this case, $k(y,s,x,t)$ is a unique solution
of the  system (5),
and solves the adjoint equation in variables $y,s$,
\cite{kal,bes1,bes2}.
Then, because $f(x),g(x)$ are locally integrable,
an immediate consequence is, \cite{watson},
that $f(x,t)$  and $g(x,t)$ are positive solutions of (5).
The identification of them with
$u(x,t)$ and $v(x,t)$ respectively, 
follows from the uniqueness of positive solutions, \cite{bes1}. 

Let us begin from a minor
generalization of (22), and define:
$${U_s(x,t)=v(x,T+s-t)\; \; , \; \; t\in [s,T]}\eqno (43)$$
Clearly, a parabolic equation (34) is satisfied by $U_s(x,t)$,
if instead
of $c(x,T-t)$, the potential $c(x,T+s-t)$ is introduced.  
An immediate propagation formula follows
$${U_s(x,t)=\int K_s(y,s,x,t)U_s(y,s)dy}\eqno (44)$$
The integral kernel $K_s$ differs from the previous $K$, (36), in the
explicit time dependence 
of the potential $c(x,T-\tau )\rightarrow c(x,T+s-\tau )$.
By putting $T=t$ in (44) we get:
$${v(x,s)=\int K_s(y,s,x,T)v(y,T)dy}\eqno (45)$$
and by the previous part of our demonstration
we know that
$${g(x,s)=\int k(x,s,y,T)v(y,T)dy}\eqno (46)$$
At this point, it is enough to prove that the identity (cf. (38))
$${K_s(y,s,x,T)=k(x,s,y,T)}\eqno (47)$$
takes place for any $s;\; 0\leq s\leq T$ .

Let us exploit the Brownian bridge scaling (13) again, so that
$${k(x,s,y,T)=[4\pi (T-s)]^{-1/2} exp[-{(x-y)^2\over {4(T-s)}}]\cdot }
\eqno (48)$$
$$\int d\mu (\alpha )  exp[-\int_s^T c({{T-\tau }\over {T-s}}x + 
{{\tau - s}\over {T-s}}y + \sqrt{T-s}\;
\alpha ({{\tau -s}\over {T-s}}),\; 
\tau )d\tau ]$$
and, analogously  
$${K_s(y,s,x,T)=[4\pi (T-s)]^{-1/2} exp[-{(x-y)^2\over {4(T-s)}}]
\cdot }\eqno (49)$$
$$\int d\mu (\alpha ) exp[-\int_s^T c({{T-\tau }\over {T-s}}y + 
{{\tau -s}\over {T-s}}x + \sqrt{T-s}\;
\alpha ({{\tau -s}\over {T-s}}),
\; T+s-\tau ) d\tau ]$$
By changing:
$${\alpha ({{\tau -s}\over {T-s}})\Rightarrow 
\alpha (1-{{\tau -s}\over {T-s}})=\alpha ({{T-\tau }\over {T-s}})}
\eqno (50)$$
and  substituting $\sigma =T+s-\tau$, where $\tau $ only is the running
variable, we finally recover
$${K_s(y,s,x,T)=[4\pi (T-s)]^{-1/2} exp[-{(x-y)^2\over {T-s}}]\cdot }
\eqno (51)$$
$$\int d\mu (\alpha )\:  exp[-\int_T^s c({{\sigma -s}
\over {T-s}}y + 
{{T-\sigma }\over {T-s}}x + \sqrt{T-s}\; 
\alpha ({{\sigma -s}\over {T-s}}),\; \sigma )(-d\sigma )]=
k(x,s,y,T)$$

Hence,
$${g(x,s)=v(x,s)}\eqno (52)$$
is valid for all time instants $0\leq s\leq T$.  This implies that
$p(y,s,x,t)=k(y,s,x,t){{v(x,t)}\over {v(y,s)}}$ defines a consistent
transition probability density of the continuous Markovian
interpolation. \\

We have succeeded to prove that:\\

(i) If  a continuous, strictly positive  Feynman-Kac kernel of the 
forward parabolic equation  
(5) is employed to solve the Schr\"{o}dinger boundary data
problem (1)  for an \it arbitrary \rm pair of nonzero probability
densities 
$\rho _0(x)$ and $\rho _T(x)$, then we can construct a Markov
stochastic  process, which is  continuous and provides for an
interpolation  between these boundary data in the time interval
$[0,T]$. 
 \\

(ii)  Given the time adjoint parabolic system (5) with bounded
solutions $u(x,t),\\  v(x,t)$ in the time interval $[0,T]$.
If the boundary densities are  defined according to (40),
then the Schr\"{o}dinger problem (1)-(3) provides us
with a  unique continuous Markov interpolation, that is compatible
with the time evolution of
$\rho (x,t)=u(x,t)v(x,t), \; t\in [0,T]$.

\subsection{Whence diffusions ?}

Our strategy,  of deducing a probabilistic solution of the
Schr\"{o}dinger boundary data problem in terms of Markov stochastic
processes running in a continuous time, was accomplished in a number
of steps accompanied by the gradual strengthening of restrictions
imposed on the Feynman-Kac potential,
to yield a continuous process (cf.
Section II.3), and eventually to get it compatible with a given
a priori parabolic evolution (Section II.4).
In a broad sense, \cite{karlin}, it can be named a diffusion.

However, this rather broad definition of the diffusion process is
significantly narrowed in the
physical  literature: while demanding the continuity of the
process, the additional restrictions are  imposed to guarrantee
that the mean and variance of the infinitesimal displacements of the
process  have the standard meaning of the drift and diffusion
coefficient, respectively, \cite{horst}.

According to the general wisdom,  diffusions arise in conjunction
with the parabolic evolution equations, since then only the
conditional averages are believed to make sense in the local
description of the
dynamics. It is not accidental that forward parabolic equations
(5) are commonly called the generalized diffusion equations. Also,
the fact that the Feynman-Kac formula involves the integration over
sample paths of the Wiener process, seems to  suggest  some diffusive
features of the Schr\"{o}dinger interpolation, even if we are unable
to establish this fact in a canonical manner.

Clearly, the conditions valid for any $\epsilon >0$: \\
(a) there holds
$lim_{t\downarrow s}{1\over
{t-s}}\int_{|y-x|>\epsilon } p(y,s,x,t)dx=0$,
(notice that (a) is a direct consequence of the stronger, Dynkin
condition, (28)),\\
(b) there exists a drift  function
$b(x,s)=lim_{t\downarrow s}{1\over {t-s}}\int_{|y-s|
\leq \epsilon }(y-x)p(x,s,y,t)dy$, \\
(c) there exists  a diffusion function
$a(x,s)=lim_{t\downarrow s}{1\over {t-s}}
\int_{|y-x|\leq \epsilon }  (y-x)^2 p(x,s,y,t)dy$,\\
are conventionally interpreted to define a diffusion process,
\cite{horst}.

To our knowledge, no
rigorous demonstration is available in the Schr\"{o}dinger problem
context, in case when the involved semigroup kernel is
\it not \rm a fundamental solution of the parabolic equation.

Let us  impose  a restriction on a lower bound of  a solution
$v(x,t)$ of the backward equation (5).
Namely, we assume that there exist constants $c_1>0,c_2>0$ such that
$v(y,s)\geq c_1 exp(-c_2y^2)$ for all $s\in [0,t], t<T$.
This property
was  found to be respected by a large class of parabolic equations,
\cite{watson2}, and it automatically ensures that the condition (25)
of Section II.3 is satisfied. Indeed:
$$0\leq lim_{|y|\rightarrow \infty }\: {1\over {v(y,s)}}
\int_{-\infty
}^{+\infty } k(y,s,x,t)v(x,t)\chi _K(x)dx \leq  $$
$$
{{1\over {c_1}} [4\pi (t-s)]^{-1/2} \: exp[M(t-s)]\cdot }\eqno (53)$$
$$ [sup_{x\in K}\:
v(x,t)]\: lim_{|y|\rightarrow \infty }\: exp(c_2y^2)\: \int_K
exp[-{(x-y)^2\over {4(t-s)}}dx] = 0 $$
if $t-s\geq \epsilon $ for sufficiently  small $\epsilon >0$
(like for example $\epsilon = 1/16c_2$). \\

It is our purpose to complete the previous analysis  by demonstrating
that, with the above assumption on $v(x,t)$, the continuous Markov
process we
have constructed actually \it is \rm the diffusion process.

Our subsequent arguments will rely on the  Dynkin treatise
\cite{dynk}.  It is well known that the infinitesimal (local)
characteristics of a  continuous Markov
process can be defined   in terms of its, so called, characteristic
operator. It is closely linked with the standard infinitesimal
(Markov) generator of the process,
and we shall take advantage of this link in below.
Let us agree, following Dynkin, to call a continuous Markov process a
diffusion, if its characteristic operator ${\cal U}$ is defined on
twice
differentiable functions (we skip more detailed definition,
\cite{dynk}). In this case $x\rightarrow x-x_0$ and
$x\rightarrow (x-x_0)^2$ allow for  the definition of a drift and
diffusion function respectively:
$${[{\cal U} (x-x_0)](x_0,s)=b(x_0,s)}\eqno (54)$$
$$[{\cal U}((x-x_0)^2)](x_0,s)=a(x_0,s)$$

By results of Sections II.3 and II.4 we know that our transition
probability density $p(y,s,x,t)=k(y,s,x,t){{v(x,t)}\over {v(y,s)}} $,
inspired by the Schr\"{o}dinger boundary data problem,
gives rise to a
continuous Markov process. To see whether it can be regarded as a
diffusion, we must verify the above two defining properties (54).

At first, let us consider the infinitesimal operator $A$ (Markov
generator) of the corresponding strongly continuous
semigroup $T_s^t: C_{\infty }(R)\rightarrow C_{\infty
}(R)$, which we have introduced via the formula (26). We are
interested in domain properties of $A$, in view of the fact that the
characteristic operator ${\cal U}$ is  a natural
extension of $A$, $A\subset {\cal U}$, \cite{dynk}.

We denote $C_c^2(R)$ the space of continuous functions with compact
support which possess continuous derivatives up to second order. For
$h\in C_c^2(R)$ we have
$${lim_{\delta \downarrow 0}\: {1\over {\delta }}[\int_{-\infty
}^{+\infty } p(y,s,x,s+\delta )h(x)dx - h(y)]=}\eqno (55)$$
$${1\over {v(y,s)}}lim_{\delta \downarrow 0}\:
{1\over {\delta }}[\int_{-\infty }^{+\infty }k(y,s,x,s+\delta )
v(x,s+\delta )h(x)dx - v(y,s)h(y)]$$

Because $v$ is continuously differentiable  with respect to time, we
have
$${v(x,s+\delta )=v(x,s)+\delta \: \partial _sv(x,s')}\eqno (56)$$
where (cf. the standard Taylor expansion formula) $s'=s+
\vartheta \delta ,\:  0\leq \vartheta \leq 1$. Hence
$${lim_{\delta \downarrow 0}{1\over {\delta }}
[\int_{-\infty }^{+\infty }
p(y,s,x,s+\delta )h(x) - h(y)]=}\eqno (57)$$
$${1\over {v(y,s)}} lim_{\delta \downarrow 0}{1\over \delta
}[\int_{-\infty }^{+\infty } dx\: k(y,s,x,s+\delta )v(x,s)h(x) -
v(y,s)h(x)] + \: $$
$${1\over {v(y,s)}} lim_{\delta \downarrow 0}[
\int_{-\infty }^{+\infty
}k(y,s,x,s+\delta )\partial _sv(x,s')h(x)dx]$$

We shall exploit the strongly continuous semigroup evolution
associated with the parabolic system (5).
Because of the domain property: $C_c^{\infty }(R)\subset D(H)$ the
smooth functions with compact support are acted upon by $H=\triangle
-c(x,s)$ and $H$ is closed as an operator on $C_{\infty }(R)$.
But then also $C_c^2(R) \subset D(H)$
and so the first term in (57) takes the form:
$${{1\over {v(y,s)}}[\triangle (vh)(y,s) - c(y,s)v(y,s)h(y)]}
\eqno (58)$$
while the second equals
$${{1\over {v(y,s)}}[\partial _sv(y,s)]f(y)= {1\over
{v(y,s)}}[-\triangle v(y,s) + c(y,s)v(y,s)]f(y)}\eqno (59)$$
Thus, (55) is point-wise convergent:
$${lim_{\delta \downarrow 0}\: {1\over \delta }[
\int_{-\infty}^{+\infty
}p(y,s,x,s+\delta )h(x)dx - h(y)]= }\eqno (60)$$
$${1\over {v(y,s)}}[(\triangle v(y,s)h(y) +
2\nabla v(y,s)\nabla h(y) +
v(y,s)\triangle h(y) - c(y,s)v(y,s)h(y) - $$
$$(\triangle v(y,s))h(y) + c(y,s)v(y,s)h(y)]= \triangle h(y) +
2({{
\nabla v}\over v})(y,s) \nabla h(y)$$

Now, we shall establish the boundedness of:
$${sup_{y\in R;0<\delta <\epsilon }\; [{1\over \delta }\:
|\int_{-\infty
}^{+\infty } p(y,s,x,s+\delta )h(x)dx - h(y)|]}\eqno (61)$$
for some  small $\epsilon $.

Because  $C_c^2(R) \subset D(H)$, so there holds
$${{1\over \delta }[\int_{-\infty }^{+\infty } k(y,s,x,s+\delta
)v(x,s)h(x)dx - v(y,s)h(y)]\rightarrow  \:
[\triangle - c(y,s)](vf)(y,s)}\eqno (62)$$
uniformly in $y$, as $\delta \rightarrow 0$. It implies that for any
compact set $K$ there is
$${sup_{y\in K;0<\delta <\epsilon }\; {1\over \delta } |
\int_{-\infty
}^{+\infty } p(y,s,x,s+\delta )h(x) - h(y)| \leq }\eqno (63)$$
$$[sup_{y\in K}\: {1\over {v(y,s)}}]\:
sup_{y\in K;0<\delta <\epsilon }\:
[{1\over \delta }|\int_{-\infty }^{+\infty }k(y,s,x,s+\delta
)v(x,s)h(x)dx - v(y,s)h(y)| + $$
$$|\int_{-\infty }^{+\infty } k(y,s,x,s+\delta )\partial
_sv(x,s')h(x)dx|] < \infty $$
We have thus the required boundedness for all $y\in K$ i.e. on compact
sets.

For $y\in R\backslash K$ we shall make the following estimations.
Because the support of $h$ is compact, we can define $supp\: h\subset
[-n,n]$ for some natural number $n$. Let  $K=[-3n,3n]$. Then:
$$sup_{y\in R\backslash K;0<\delta <\epsilon }\: {1\over \delta
}|\int_{-\infty }^{+\infty } p(y,s,x,s+\delta )h(x)dx - h(y)|= $$
$${sup_{y\in R\backslash K;0<\delta <\epsilon }\: {1\over \delta }|
\int_K
p(y,s,x,s+\delta )h(x)dx| \leq }\eqno (64)$$
$$[sup_{x\in K}\: |h(x)|]\: sup_{y\in R\backslash K;0<\delta
<\epsilon }\:
{1\over \delta }{1\over {v(y,s)}} \int_K k(y,s,x,s+\delta  )
v(x,s+\delta )dx \leq $$
$$[sup_{x\in K}\: |h(x)|]\: [sup_{x\in K;s\leq s'\leq s+\epsilon }\:
v(x,s')]\: sup_{y\in R\backslash K;0<\delta <\epsilon } \: {1\over
\delta
}{1\over {v(y,s)}}\int k(y,s,x,s+\delta )dx$$

In view of our assumption $v(y,s)\geq c_1exp(-c_2y^2)$,
there holds:
$${sup_{y\in R\backslash K;0<\delta <\epsilon }\:
{1\over \delta }|\int
p(y,s,x,s+\delta )h(x)dx| \leq }\eqno (65)$$
$$C\cdot sup_{|y|\geq 3n;0<\delta <\epsilon
}\; exp(c_2y^2)\: \delta ^{-3/2}\: \int_{-n}^{+n}
exp[-{{(x-y)^2}\over {4\delta }}\: dx$$
where
$${C=c_1(4\pi )^{-1/2}exp(M\epsilon )\: [sup_{x\in K}|h(x)|]\:
sup_{x\in
K;s<s'<s+\epsilon }v(x,s')}\eqno (66)$$
If we choose $\epsilon =1/16c_2$, then
$${exp(c_2y^2)\: \int exp[-{{(x-y)^2}\over {4\delta }}]\: dx
\leq 4\delta
exp(-{n^2\over \delta })}\eqno (67)$$
for every $|y|\geq 3n$, and so
$${sup_{y\in R\backslash K;0<\delta <\epsilon }\:
{1\over \delta }|\int
p(y,s,x,s+\delta )h(x)dx|\leq 4Csup_{0<\delta <\epsilon }\delta
^{-1/2}exp(-{n^2\over \delta }) < \infty }\eqno (68)$$
Consequently, the desired boundedness (62) holds true for all
$y\in R$,
together with the previously established point-wise convergence (61).

Altogether, it
means, \cite{dynk}, that the weak generator of $T_s^t$ is defined at
least on $C_c^2(R)$.
Moreover, while acting on $h\in C_c^2(R)$ it gives
$\triangle h + (\nabla ln \: v)\nabla h$.
Because $T_s^t$ is strongly continuous in $C_{\infty }(R)$, the
Markov generator $A$ coincides with the weak generator, \cite{dynk},
i.e. $A=\triangle +(\nabla ln\: v)\nabla $  on $C_c^2(R)$.

Finally,  let us choose $h_0\in C_c^2(R)$ such that $h_0(x)=1$ in some
neighbourhood of the point $x_0$. Then, $(x-x_0)h_0(x)$ and
$(x-x_0)^2h_0(x)$ both belong to $C_c^2(R)$ and therefore:
$${A[(x-x_0)h_0 ](x_0,s) = \triangle [(x-x_0)h_0 ](x_0) + }
\eqno (69)$$
$$2(\nabla ln\: v)(x_0,s)\nabla  [(x-x_0)h_0](x_0)=2(\nabla ln\:
v)(x_0,s)\:$$
$$ A[(x-x_0)^2h_0](x_0,s) = 2 $$

Because $A\subset {\cal U}$ and ${\cal U}$ is a local
operator,\cite{dynk}, we have  the following inclusion $C_c^2(R)
\subset
D({\cal U})$ and (we can get rid of $h_0$):
$${[{\cal U}(x-x_0)](x_0,s)=2(\nabla ln\: v)(x_0,s)}\eqno (70)$$
$$[{\cal U}(x-x_0)^2](x_0,s)=2$$

It means that we indeed obtain a diffusion process with the drift
$\nabla ln\: v$ and a constant diffusion coefficient, according to
the standards of \cite{zambr,nel,carlen}.

It is worth emphasizing that since
  $(x-x_0)h_0(x)$ and $(x-x_0)^2h_0(x)$
belong to $D(A)$, and since functions from $C_c^2(R)$ can be used to
 approximate,
under an integral, an indicator function of the set
$[x_0-\epsilon,x_0+\epsilon ],\epsilon >0$, we can directly evaluate:
$${lim_{t\downarrow s} {1\over {t-s}}\int_{-\infty }^{+\infty }
p(x_0,s,x,t)(x-x_0)h_0(x)dx = }\eqno (71)$$
$$lim_{t\downarrow s} {1\over {t-s}}\int_{|x-x_0|\leq \epsilon }
p(x_0,s,x,t)(x-x_0)dx = 2(\nabla ln\: v)(x_0,s)$$
and similarly
$${lim_{t\downarrow s}{1\over {t-s}}\int_{|x-x_0|\leq \epsilon
}p(x_0,s,x,t)(x-x_0)^2dx = 2}\eqno (72)$$

Because the Dynkin condition (28) implies that
$${lim_{t\downarrow s} {1\over {t-s}}\int_{|x-x_0|> \epsilon }
p(x_0,s,x,t)dx = 0 }\eqno (73)$$
we arrive at the commonly accepted definition of the diffusion
process,
summarized in formulas (71)-(73), with the functional expression for
the drift, (71), given in the familiar, \cite{zambr,nel,blanch},
gradient form.

\section{Nonstationary Schr\"{o}dinger dynamics: from the Feynman-Kac
kernel to diffusion process}

In our  previous paper \cite{olk}, the major conclusion was that in 
order to give 
a definitive probabilistic description of the quantum dynamics as a 
\it unique \rm diffusion process solving Schr\"{o}dinger's
interpolation problem, a suitable Feynman-Kac semigroup must be
singled out. Let us
point out  that the measure preserving dynamics, permitted in the 
presence of conservative force fields, was investigated in
\cite{blanch}, see also \cite{carm,freid}.

The present analysis 
was performed quite generally and extends to the dynamics affected by 
time dependent external potentials, with no 
clear-cut discrimination between the nonequilibrium statistical
physics  and essentially quantum  evolutions.
The formalism of Section II encompasses both groups of problems.
Presently, we shall  restrict our   discussion
 to the free Schr\"{o}dinger picture quantum dynamics. Following
Ref. \cite{olk} we shall discuss the rescaled problem so as to 
eliminate all dimensional constants.

The free Schr\"{o}dinger evolution $i\partial _t\psi =
-\triangle \psi $ 
implies  the following propagation of a specific Gaussian
wave packet:
$${\psi (x,0)=(2\pi )^{-1/4} exp\: (-{{x^2}\over {4}})\; 
\longrightarrow  }\eqno (74) $$
$$\psi (x,t)=({2\over \pi })^{1/4} \; (2+ 2it)^{-1/2} 
exp[-{x^2\over {4(1+it)}}]$$
So that 
$${\rho _0(x)=|\psi (x,0)|^2=(2\pi )^{-1/2}\: exp[-{x^2\over 2}] 
\longrightarrow }\eqno (75)$$
$$\rho (x,t)=|\psi (x,t)|^2= [2\pi (1+t^2)]^{-1/2}\: 
exp [-{x^2\over {2(1+t^2)}}]$$ 
and the Fokker-Planck equation (easily derivable from the standard 
continuity equation $\partial _t\rho =-\nabla (v\rho ),\; v(x,t)=
xt/(1+t^2)$) holds true:
$${\partial _t\rho = \triangle \rho - \nabla (b\rho )\; \;  , \; \; 
b(x,t)= - {{1-t}\over {1+t^2}}\: x} \eqno (76)$$

The Madelung factorization $\psi =exp(R+iS)$ implies (notice that 
$v=2\nabla S$ and $b=2\nabla (R+S)$) that the 
related real functions  $\theta(x,t)=exp[R(x,t)+S(x,t)]$ and
$\theta _*(x,t)=
exp[R(x,t)-S(x,t)]$ read:
$$\theta (x,t)=[2\pi (1+t^2)]^{-1/4} exp(-{x^2\over 4}\: 
{{1-t}\over {1+t^2}} - {1\over 2} arctan\: t)$$
$${\theta _*(x,t)=[2\pi (1+t^2)]^{-1/4} exp(-{x^2\over 4}\: 
{{1+t}\over {1+t^2}} + {1\over 2} arctan\: t)}\eqno (77)$$
They solve a suitable version  of the general parabolic equations (5), 
namely :
$${\partial _t \theta =-\triangle \theta + c  \theta }
\eqno (78)$$
$$\partial _t\theta _* =\triangle \theta _* - c  \theta  _*$$
with 
$${c(x,t) = {x^2\over {2(1+t^2)^2}} -
{1\over {1+t^2}} = 
2{{\triangle \rho ^{1/2}}\over {\rho ^{1/2}}}}\eqno (79)$$

Anticipating further discussion, let us mention that the Feynman-Kac
kernel, in this case, \it is \rm a  fundamental solution of the
time adjoint system (78).
For clarity of exposition, let us recall that a \it fundamental
solution \rm of the forward parabolic equation (5) is a continuous
function $k(y,s,x,t)$,
defined for all $x,y,\in R$ and all $0\leq s<t\leq T$, which has the 
following two properties: \\
(a) for any fixed $(y,s)\in R\times (0,T)$, the function
$(x,t)\rightarrow k(y,s,x,t)$ is a regular (i.e. continuous and 
continuously differentiable the needed number of times) solution of
the  forward equation (5) in $R\times (s,T]$\\
(b) for all continuous functions $\phi (x)$ with a compact support, 
there holds $lim_{(t,x)\rightarrow (s,z)}$
$\int_{-\infty }^{+\infty } 
k(y,s,x,t)\phi (y) dy = \phi (z)$.\\ 

First, we need to verify (this will be done self-explanatorily) that
$c(x,t)$, (79), is H\"{o}lder
continuous of exponent one on every compact subset of $R\times [0,T]$. 
It follows from direct estimates:
$${|c(x_2,t_2)-c(x_1,t_1)|\leq {1\over 2} |{x_2^2\over {(1+t_2^2)^2}}-
{x_1^2\over {(1+t_1^2)^2}}|\: +\:
|{1\over {1+t_2^2}}-{1\over {1+t_1^2}}|\leq }
\eqno (80)$$
$${1\over 2}|{x_2\over {1+t_2^2}}-{x_1\over {1+t_1^2}}|(
{|x_2|\over {1+t_2^2}} + {|x_1|\over {1+t_1^2}})\: + \: |t_2 -t_1|\:
{{|t_1|+|t_2|}\over {(1+t_1^2)(1+t_2^2)}}$$
But, in case of $|x_1|,|x_2|\leq K$ and $|t_1|,|t_2|\leq T$ we have
$${|c(x_2,t_2)-c(x_1,t_1)|\leq K\: |{x_2\over {1+t_2^2}}- {x_1\over
{1+t_1^2}}|\: +\: 2T|t_2-t_1|}\eqno (81)$$
Furthermore:
$${|{x_2\over {1+t_2^2}}-{x_1\over {1+t_1^2}}|\leq |{{x_2-x_1}\over
{(1+t_2^2)(1+t_1^2)}}|\: +\: |{{x_2t_1^2-x_1t_2^2}\over
{(1+t_2^2)(1+t_1^2)}}|\leq }\eqno (82)$$
$$|x_2-x_1|+T^2|x_2-x_1|+2KT|t_2-t_1|$$
implies (the new constant $C$ majorizes all remaining ones)
$${|c(x_2,t_2)-c(x_1,t_1)|\leq C\: (|x_2-x_1|\: +\: |t_2-t_1|)\leq
\sqrt
2\: C\: [(x_2-x_1)^2+(t_2-t_1)^2]^{1/2}}\eqno (83)$$

Let us also notice that we can introduce an auxiliary function
$h(x,t)=arctan\: t$ such that there holds
$${\triangle h-c(x,t)h-\partial_t
h=-{{x^2h(x,t)}\over {2(1+t^2)^2}}\: \leq \: 0}\eqno (84)$$
We have thus satisfied the crucial assumptions I and II of Ref.
\cite{bes2}.
As a consequence, we have granted the existence of a fundamental
solution $k(y,s,x,t)\geq 0$.
Moreover, for every bounded and continuous function $\phi (x),
\: |\phi
(x)|\leq C$, where $C>0$ is arbitrary, the function
$${u(x,t)=\int_{-\infty }^{+\infty } k(y,0,x,t)\phi (y)dy}
\eqno (85)$$
is a solution of the Cauchy problem,
i.e. solves (79) under the initial
condition $u(x,0)=\phi (x)$, so that $|u(x,t)|\leq C$.
All that implies the uniqueness of the fundamental solution
$k(y,s,x,t)$, and in view of $-c(x,t)\leq 1$ its strict positivity.
The function $k(y,s,x,t)$ is also a  solution of the adjoint
equation with respect to variables $y,s$:
$\partial _sk=-\triangle _yk + c(y,s)k$ in $R\times [0,T)$.
It is obvious that the Chapman-Kolmogorov composition rule holds true,
in view of the validity of the Feynman-Kac representation in the
present case.\\
Basically, we must  be satisfied with the Feynman-Kac representation
 of the fundamental solution,
whose existence we have granted so far.
In our case, the so called parametrix method,
\cite{kal}, can be  used to construct fundamental solutions.
In fact, since $c(x,t)$ is
locally Lipschitz i.e. H\"{o}lder continuous of exponent one and
quadratically bounded $|c(x,t)|\leq x^2+1$, the infinite series:
$${k(y,s,x,t)=\: \sum_{n=0}^{+\infty } (-1)^n  k_n(y,s,x,t)}
\eqno (86)$$
where $k_0(y,s,x,t)=[4\pi (t-s)]^{-1/2}
\: exp[-(x-y)^2/4(t-s)]$ is the heat kernel  and
$${k_n(y,s,x,t)=\int_s^td\tau \int _{-\infty }^{+\infty } dz\:
c(z,\tau
)\: k_{n-1}(y,s,z,\tau )k_0(z,\tau , x,t)}\eqno (87)$$
are known to converge for all
$x,y\in R$, $0\leq s<t\leq T$, and $t-s<T_0$ where $T_0<T$, and
define the fundamental solution, \cite{krzyz}.

By  putting $p(y,s,x,t)=k(y,s,x,t){{\theta (x,t)}
\over {\theta (y,s)}}$ we arrive at the fundamental solution of
the second Kolmogorov (Fokker-Planck) equation
$${\partial _tp(y,s,x,t) =\triangle _xp(y,s,x,t) -\nabla _x
[b(x,t)p(y,s,x,t)]}\eqno (88)$$
where  $b=2{{\nabla \theta }\over {\theta }}$ and
$\rho =uv$, and in particular $\rho =\theta \theta _*=|\psi |^2$,
are  consistently propagated by $p$.
It is the transition probability density of the
Nelson diffusion associated with the solution (74) of the
Schr\"{o}dinger equation, and at the same time a solution of the first
Kolmogorov (backward diffusion) equation
$${\partial _sp(y,s,x,t)=-\triangle _yp(y,s,x,t) -
b(y,s)\nabla _yp(y,s,x,t)}\eqno (89)$$
Equations (88), (89) prove that the pertinent process is a
diffusion: it has the standard local (infinitesimal) characteristics
of the  diffusion process, \cite{horst}.

Obviously, the above definition of $p$ in terms of $k$ induces the
validity of  the compatibility condition
$${c(x,t)=2[\partial _tln\: \theta(x,t)  + {1\over 2}[{{b^2(x,t)}\over
2}+\nabla b(x,t)]}\eqno (90)$$
connecting the drift  of the diffusion process with the
Feynman-Kac potential  governing its local dynamics: cf. Refs.
\cite{carm,blanch} and \cite{garb2} where the Ehrenfest theorem
analogue was formulated for general (non-quantal included)
Markovian diffusions.

Let us point out that our  quantally motivated example  was chosen
not to show up a typical for quantum wave functions property of
vanishing somewhere.
 In fact, because of restricting our considerations to strictly
positive Feynman-Kac kernels and emphasizing the uniqueness
of solutions, we  have left aside an
important group of topics pertaining
to solution of the Schr\"{o}dinger boundary data problem when:\\
(i)  the boundary densities have zeros\\
(ii) the interpolation itself is capable of producing zeros of the
probability density, even if the boundary ones have none.\\
Only the case (i) can be (locally) addressed by means of
strictly positive
semigroup kernels, however the uniqueness of solution is generally
lost
in space dimension higher than one, \cite{fort,beur,jam}.
General existence theorems are available \cite{carlen,carm} and
indicate that one deals with diffusion-type processes in this case,
see e.g also \cite{zambr,zambr1,blanch}.\\
The case (ii) seems not to be ever considered in the literature, see
however \cite{garb1}.
\vskip0.2cm
{\bf Acknowledgement}: Both authors receive a financial support from 
the  KBN research grant No 2 P302 057 07. We would like to thank 
Professors John Klauder and Gert Roepstorff for correspondence
concerning the differentiability of Feynman-Kac kernels and
Jean-Claude Zambrini for reference suggestion.

\end{document}